# Analisys of women leadership in enterprise social networks

Giorgia Di Tommaso, Giovanni Stilo and Paola Velardi
Dipartimento di Informatica
Sapienza Università di Roma
Italia
$\{stilo, velardi\}@di.uniroma1.it$

*Abstract*—**This paper describes a Social Network Analysis toolkit to monitor an Enterprise Social Network and help analyzing informal leadership as a function of social ties and topic discussions. The toolkit has been developed in the context of a regional project, Fiordaliso, funded by Regione Lazio (a region of central Italy) and leaded by Reply, an international network of specialized companies in the field of digital services.**

*Social network analysis; women leadership; LeadRank; topic clustering; informal leadership; enterprise social networks*

## I. INTRODUCTION

Fiordaliso is a project financed in late 2014 by Regione Lazio (a region of central Italy) in the context of a local government funding program[1] to encourage the sharing of knowledge and innovation capacity of enterprises and make them more competitive on the international market.

One of the objectives of *Fiordaliso* is to develop a social network analysis layer on the top of an enterprise social platform, TamTamy [2]. TamTamy has been designed by Reply, an international network of specialized companies in the field of digital services, with the purpose of increasing engagement, brand awareness and knowledge sharing among the employees of public and private organizations. More specifically, *Fiordaliso* is aimed at analyzing the emergence of the so-called "informal leadership" in organizations. According to Shaman Grimsley [1] informal leadership is *"the ability of a person to influence the behavior of others by means other than formal authority conferred by the organization through its rules and procedures"*. In the context of our project, "other" is the environment provided by an on-line enterprise social network: we refer to this specific kind of informal leadership as to *network leadership*.

A second objective of the project is to detect and favor the emergence of women leadership in organizations. Several studies encompassing both experimental and survey data have shown that social networks represent a conductive environment for the emergence of women attitude towards informal leadership. For example, in [2] the authors suggest that women's social orientation is well suited to achieving a central role in social networks and therefore, the emergence of their leadership. Similarly, Eagly and Karau [4] argue that women can reach higher leadership levels with respect to men, when teamwork implies more social interaction skills. Finally, in [3] it is experimentally shown that, when networks are perceived to be cohesive (i.e. when there are tight relationships among individuals) women are seen more charismatic than men leaders.

In order to meet these objectives, *Fiordaliso* was first targeted at identifying a number of suitable leadership measures, based both on content and structural features of the underlying social platform, and then, at designing data mining and visualization tools in order to a) provide evidence of emerging leadership behavior, and b) compare these findings with the actual role and gender of an organization's employees. This paper describes the leadership measures and the SNA layer of the platform, and provides preliminary data analysis concerning issue a). The paper is organized as follows: Section II presents related work in the area of leadership analysis in on-line social networks. Section III describes the TamTamy network model and the leadership rank model adopted in our study. Section IV summarizes the *Fiordaliso* toolkit features and provides a preliminary analysis of data extracted from a 3-year anonymized dataset of users' threads. Finally, Section V is dedicated to concluding remarks.

---

[1] http://www.regione.lazio.it/europaimprese/innovazione-reti-d-impresa/insieme-per-vincere.php
[2] http://www.reply.eu/tamtamy-reply/en/

## II. PROBLEM FORMULATION AND RELATED WORK

Sociological and management literature has paid a great attention to the problem of characterizing leadership models in organizations. If we restrict to informal leadership, the qualities of an informal leader have been characterized as follows [5]:

- Competence (*confidence, ability, knowledge, willingness, example, influence*);
- Organizational culture (*encouragement, ideas, asking, opportunities*);
- Situational requirements (*ability, organization, goals, effectiveness, company, team*).

As further stated in [6], network leadership is more about influence than control, requiring leaders to create a work environment based on *autonomy, empowerment, trust, sharing,* and *collaboration*. In particular, empowerment is considered to be an important leadership quality in communities [7], in order to foster a greater responsibility of employees through a knowledge sharing and participation in decision processes and problem solving. More precisely, empowerment[3] is defined as the management practice of sharing information, rewards and power with the employees. In our project we faced with the problem of formalizing these qualities in order to produce a quantitative leadership ranking in on-line enterprise social network. To the best of our knowledge, this is the first study providing a quantitative analysis of gender leadership in on-line enterprise social networks. There are however three lines of research closely related with our work. The bulk of research is concerned with the definition of leadership in generalized social networks. The majority of scholars in this field define some variant of the Page Rank algorithm [8]. In [9] the authors propose an algorithm, named Dynamic Opinion Rank, which is based on Page Rank and content analysis. Users are classified on the basis of their expertise on the discussion topics and on the comments (positive or negative) of other users. An "influence degree" is then assigned to each message. Influence is propagated between authors using Page Rank. In [10] another variation of PageRank is proposed, named *LeaderRank*. As in the previous work, the notion of leadership is tightly connected with that of competence, but rather than modeling competence as a function of message content, the authors use bookmarking, which is available in several discussion networks (in the paper, experiments are conducted on *Delicious*). Khorasgan et al. [11] introduce *TopLeaders*, an algorithm that, first, identifies clusters of connected components using the k-means algorithm. Then, within each cluster, a team leader is identified. A number of other papers (e.g. [12]) identify leaders with reference to their centrality within communities.

A second line of research analyzes the dynamics and strength of relations in enterprise social networks. A notable study (among the few) in this area is [13]. Based on a 6 months dataset collected from a large enterprise social network, the authors analyze the influence on the user graph connectivity of several factors such as hierarchical relations (co-worker, supervisor, etc.), distance between headquarters of communicating users, and other features. In order to quantify the relations between influencing features and network dynamics they use a logistic regression model. The results show that both users' location and hierarchical differences do influence network statistics such as the number of influential users, the likelihood of interaction between user pairs and the graph connectivity. Only one study has analyzed users' behavior in social networks according to gender [14]. The study presents a multiplex network model to analyses users' behavior in a social network representing users' interaction in an on-line game. The multiplex network models 6 types of relationships (friendship, enmity, etc.) between two types of users (male, female). Experimental results show that women cumulate a significantly higher number of credits in the game, but they exhibit a lesser tendency to risk. Furthermore, they perform a higher number of positive actions and, as a consequence, they attract more positive actions in return. As in many other sociological studies, women are found to be more communicative, but less prone to homophily than men.

## III. MEASURING INFORMAL LEADERSHIP

This Section describes the methods and algorithms used to analyze an enterprise social network and the behavior of its users. First, we describe the structure of the TamTamy entreprise network. Though the type of data that can be derived from a social network obviously depend on the specific platform, many of the features of TamTamy are indeed common to other popular enterprise social newtorks, like e.g Jive[4]. Basically, these environments provide a collaboration space to share knowledge and documents, communicate with co-workers, search and find relevant content and people within a company.

---

[3] http://www.businessdictionary.com/definition/empowerment.html

[4] https://www.jivesoftware.com/products-solutions/jive-n/

## A. The TamTamy network

Though the TamTamy platform provides a number of services, we are concerned here mainly with social networking and communication. Messages are started by a thread initiator (the *author*). The other users (*commenters*) are then free to contribute and to rate messages with like and dislike, much in the Facebook style. The basic format of each thread is the following:

content_id,content_title,content_description,publication_date,content_t ags,content_author,user_title,gender(0=male,1=female),{comment_id, comment_description,comment_date,comment_author,commenting_us er_title,gender(0=male, 1 = female)}$^n$

The *content_tags* are free, although there exist a number of pre-defined tags. *Title* refers to the author/commenter role in the company (manager, director, consultant, senior consultant, partner, external). A thread excerpt (amended for privacy reasons) is shown in Figure 1. Note that only rarely is the recipient explicitly indicated in the message. Threads are in English, German and Italian, and often the text includes a mix-up of languages since many technical terms are in English even when the conversation is in another language. Moreover, the use of mixed terms (e.g. *pagamento online*) is quite common. This "mixed linguality" prevents from using even basic natural language processing tools such as part of speech tagging, as discussed later.

## B. Modeling network connectivity

We model the network as a three-way multiplex network. The first layer $G_1(N, E_1)$ models the activity of authors, the second layer $G_2(N, E_2)$ models the activity of commenters, and the third layer $G_3(N, E_3)$ that of raters. Figure 2 shows co-workers activities and roles within the network. In the first layer $G_1(N, E_1)$, for any thread $\theta$ initiated by any author $i$ we add an edge $e_1(i,j)$ with $w_\theta(i \rightarrow j) = 1$, whenever $j$ posts a comment in $\theta$. This models the fact that author $i$ has empowered user $j$. The *cumulative empowering weight* of an edge $e_1(i,j)$ is computed as:

$$w_e(i \rightarrow j) = \frac{\sum_{\forall \theta} w_\theta(i \rightarrow j)}{\sum_{\forall \theta, k} w_\theta(k \rightarrow j)} \quad (1)$$

The adjacency matrix $E$ (*empowerment matrix*) associated with $G_1$ is left stochastic, with each column summing to 1. Edges $e_2(i,j)$ in $G_2(N, E_2)$ are created whenever a user $i$ posts a comment for the first time in a thread $\theta$ authored by $j$. Subsequent answers update the weight of $e_2(i,j)$. Given the sequential thread structure, comments are considered as implicitly directed to the author, unless the message includes the name of another recipient in the thread (as in Figure 1). Note that authors themselves can add comments during a discussion, and in this case, they are treated as commenters. Any comment $h$ in a thread $\theta$ has a weight computed as follows:

$$w_h(i \rightarrow j) = (0.5 + 0.5/k)$$

where $k$ is the sequential order of the comment in the thread ($k=1$ if a user is the first who answers). This models the idea that the first commenter is more competent, or more willing to collaborate, than others (alternatively, $k$ can be set to represent the difference between the time stamp of the author's message and that of the commenter). The *cumulative collaboration weight* of an edge $e_2(i,j)$ sums all answers provided by user $i$ to user $j$ in any thread within a considered time span $W$:

$$w_c(i \rightarrow j) = \frac{\sum_{\forall h, \theta} w_h(i \rightarrow j)}{\sum_{\forall h, \theta, k} w_h(k \rightarrow j)} \quad (2)$$

---

```
Thread id:115863
Title:" ** - 2014 ** Page "
Author:b80a4fcb00acc
Timestamp:"2014-01-02 13:41:20"
Role: Consultant
Tags:client+requirements , home+page , km42 , mockup
" Hi  @ddd22ccb attached a first proposal regarding the **
Page related to the feedback we shared by phone. I would
share also with the team to have feedback regarding  [..].
It is very essential is that ok for you? After you
feedback I will ask to ** support to evaluate the
technical effort required. Thanks "

Commenter: ddd22ccb
Commenter Role: Director
timestamp:"2014-02-27 17:51:10"
" ok , let ' s proceed "
[…]
```

Figure 1. Example of an (amended) thread

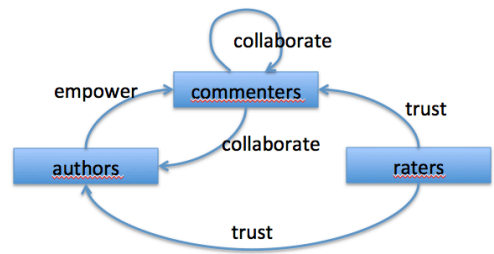

Figure 2. Co-workers roles and actions in the enterprise social network

The collaboration matrix $C$ is, as for $E$, left stochastic. The third layer $G_3(N, E_3)$ models the rating activity of users. Edges $e_3(i,j)$ are weighted with the trust $w_t(i \rightarrow j)$ of rater $i$ in user $j$. We first introduce the quantity:

$$trust(i \rightarrow j) = 0.5 + 0.5 \times \sum_{h \in m(j)} \delta_h(i \rightarrow j) \quad (3)$$

where $m(j)$ is the set of messages generated by user $j$ (either as an author or as a commenter), $h$ is a message in $m(j)$ and $\delta_h(i \to j) = +1$ if $i$ likes $h$, -1 if $i$ dislikes, and 0 if no opinion is expressed. According with (3), $0 \leq trust < 0.5$ indicates distrust. We then define:

$$w_t(i \to j) = trust(i \to j) / \sum_k trust(i \to k) \quad (4)$$

The $T$ matrix, named the *credibility matrix*, is right stochastic. We further denote with $\mathcal{M}$ the 3-way $N \times N \times 3$ tensor of the multiplex network. The third dimension of the network represents the <u>leadership qualities</u> of users: *empowerment, collaboration* and *credibility*. In fact, a user $i$ with high empowerment capability ( $\sum_j w_e(i \to j)$) potentially meets the organizational and situational requirements [5,6] since he/she started many threads (*encouragement, ideas, asking, opportunities*) and received many answers (*effectiveness, company, team*). Similarly, high collaboration attitude ($\sum_j w_c(i \to j)$) and credibility $\sum_j w_t(j \to i)$) for a user $i$ are a proxy of the competence requirement (*confidence, ability, knowledge, willingness, example, influence*).

A simple assumption would be to compute for every user the engagement, collaboration and credibility ranks $r^e, r^c$ and $r^t$ using any centrality measure, for example, Page Rank [8], and then computing some heuristic function to combine these indicators. However it is assumed that the centrality of each node in one layer affects the centrality of the same node in any other layer. Therefore, a better measure is *Multiple Page Rank (MPR)*, introduced in [15]. First, we note that PageRank centrality depends on the in-degree of nodes, while according to our formulation, *empowerment* and *collaboration* of a node depend on the weight of outgoing edges. Therefore, we need to invert the direction of edges in the corresponding graphs. We denote with $E^T, C^T$ and $T$ the slices of an $\mathcal{M}'$ tensor, and with $w_{e^T}(i \to j) = w_e(j \to i)$, $w_{c^T}(i \to j) = w_c(j \to i)$ and $w_t(i \to j)$ the corresponding matrix cells. MPR is iteratively defined in what follows. At the first layer (*empowerment*) we initially have:

$$r_i^e = \alpha^e \sum_j w_{e^T}(i \to j) r_j^e + (1 - \alpha^e) \frac{1}{N}$$

which corresponds to the monoplex PageRank formulation with teleporting. We then include the information about the other layers, and obtain the general MPR formulation:

$$r_i^k = \alpha^k \sum_j (r_i^{k-1})^\beta w_k(i \to j) r_j^k + (1 - \alpha^k) (r_i^{k-1})^\gamma / \langle (r^{k-1})^\gamma \rangle N \quad (5)$$

where in our case $k = 1,2,3$ and the symbol $\langle ... \rangle$ indicates the average. In this formulation, the teleporting factor $\alpha^k$ is layer-dependent, and the exponents $\beta, \gamma \leq 1$ are set to tune the influence of one layer on the other. Since in a multiplex network layers are not connected (contrary to interconnected multilayer networks), teleporting does not allow to jump from one node of a layer to another node of another layer, though it is influenced by the previous layer, as shown in formula (5). To efficiently calculate stationary values in MPR, an iterative "*divide et impera*" strategy is adopted[5] in line with [15] and [17].

*C. Modelling communications*

Collaboration, empowerment and credibility of users may vary both in time and, mainly, according to the specific topics being discussed. A user may be highly credible when he/she discusses about, e.g. *mobile apps*, and be much less confident on *business models*. Consequently, informal leadership should be analyzed also in the context of specific topics. The content of threads, besides their multilinguality and mixed-linguality, greatly differ also in the type of discussed topics. A few threads are on leisure topics (for example, the organization of a football match) however the majority is on technical or administrative topics. We perform topic summarization in two steps: first, by extracting relevant terminology from threads, second, by generating clusters of co-occurring terms. A commonly used approach in literature is to use stemmed words as terms and a latent topic model (e.g. LDA [18] or one of its many variants). This solution turned out to perform poorly due both to mixed linguality and to the reduced dimension of messages. To extract more significant terms, first, we index only "content" tokens, or *concepts*, identified as those words mapping with BabelNet [19], a freely available semantic network covering more than 50 languages and more than 13 millions concepts. In this way, the specific language in which a concept is expressed does not matter. Then, we extract concept n-grams that are either consecutive (i.e. compounds) or separated by prepositions and determiners. Finally, we extract *concept cliques* using the Bron-Kerbosch clique detection algorithm as described in [20], with the restriction that each element in the clique exceeds an experimentally defined frequency threshold $m$. An example of multi-mixed-lingual topic is the following:

> *investimenti_online*
> *digital_marketing*
> *perspective_engage*
> *analisi_delle_performance*

Note in the example the presence of several "mixed language" concepts (e.g. *investimenti online, analisi delle performance*). This is rather common in work environments where the use of English technical terms

---

[5] Details are omitted for brevity, interested readers refer to [15] and [17].

dominates. Topics are extracted within temporal windows of length *W* (we experimented with *W*=1 week or 1 month). Cosin-similarity is used to cluster topics vertically (within the same *W*) and horizontally along the temporal line, in order to generate *topic streams* $s(\tau)$. An example of two topics assigned to the same stream is the following:

TOPIC#:169
user_experience
news_pay
carte_di_credito
carta_di_credito
metodo_di_pagamento
american_express
pagamenti_online

TOPIC#:162
mobile_pos
pay_reply
user_experience
soluzione_di_mobile_pos
sistemi_di_pagamento
circuiti_di_pagamento
gestione_coupon

For the sake of space we do not compare in detail our algorithm with LDA, however we mention that topics extracted with different methods have been comparatively evaluated by company experts who found the solution proposed here significantly more performant in terms of quality than, e.g. using LDA.

Given a topic stream $s(\tau)$ within a temporal windows *W*, we are then able to generate the network of users participating in $\tau$, and to derive all measures described in previous section with reference to $\tau$. This is particularly relevant as far as *competence* and *credibility* are concerned, since both ranks may depend on specific discussion topics.

IV. THE FIORDALISO DASHBOARD

This Section shortly describes some of the features of the Fiordaliso dashboard and presents a summary of experimental results. We experimented our algorithms and toolkit on a 3-year dump (2012-2014) of anonymized and partly amended (i.e. without attachments) TamTamy threads mostly in English and Italian, with a few German threads. Parameters in (5) are all set to 1. Overall, we analyzed over 100.000 threads from more than 2000 different users. Active female users have been between 23% and 24% in years 2012, 2013 and 2014. The dashboard provides a large number of statistics, trend graphs, topic graphs and streams, users' and global rankings that we cannot describe in detail for the sake of space. We only provide here a summary of main findings. The most relevant rankings computed for individual users are the previously described leadership rank and brokerage[6] rank, a measure of the ability of individuals to connect communities. Figure 4 shows that the probability mass of top ranked women leaders and brokers are significantly higher that the prior probability of women (the thick line in Fig.4). Negative peaks correspond to holidays (typically, on august), with a persistent negative trend during august-november 2014, due to a change in internal relationships of the company.

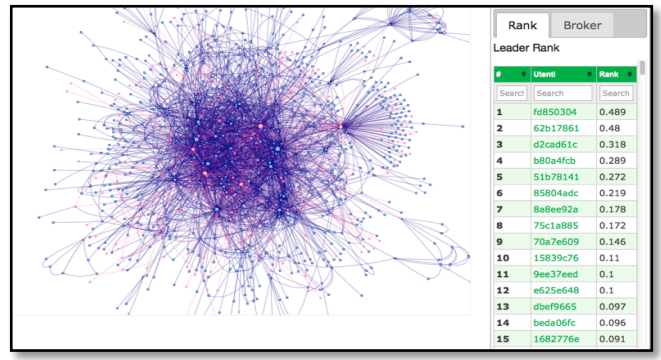

Figure 3. Snapshot of TamTamy social network during 6 months since January 2014. Pink circles are women, blue squares are men. Leadership ranks are shown on the right side. First ranked is a woman Senior Consultant.

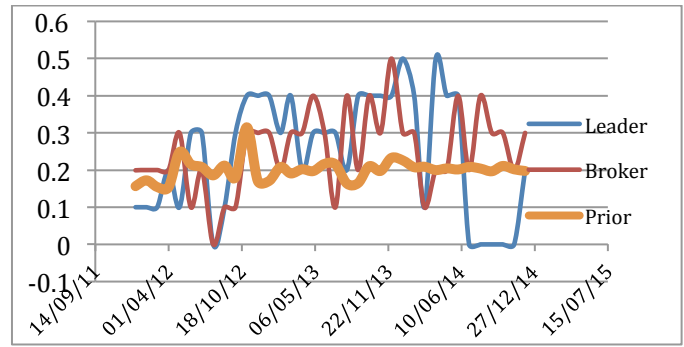

Figure 4. Women Leadership and brokerage ranks compared with prior probability of women along the 3 years timeline.

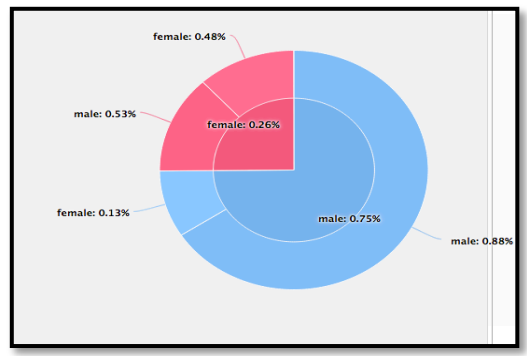

Figure 5. Homophily measured as the tendence of women/men to answer women's (men's) posts

Figure 5 shows another interesting finding: contrary to what claimed in [14], our data consistently show during the three years that women have a higher tendency to homophily than men (the prior probability of a women to answer a women equals the prior probability of women authors, which is around 25%, while Figure 5 shows that 48% of women respond to women, which is statistically very significant). On the other side, we remark that experiments in [14] are conducted on the social network of an on-line game, an environment that is clearly less formal than an enterprise social network. Similarly to [13] we found that roles do influence social relations: co-workers tend to answer more rapidly and frequently to managers, however, this is less evident when a message is started by a female manager, which shows that, unfortunately, the authority of males is acknowledged more than that of female leaders.

---
[6] http://faculty.ucr.edu/~hanneman/nettext/C9_Ego_networks.html#brokerage

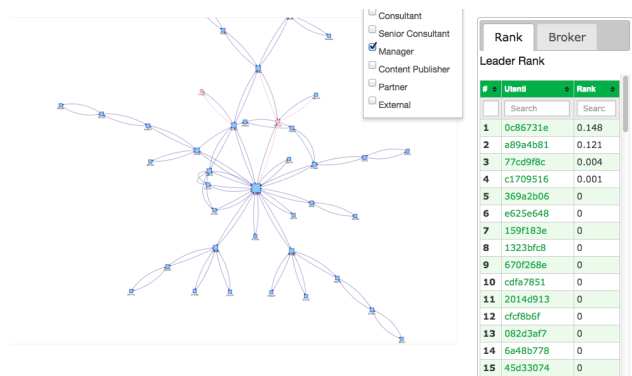

Figure 6. Social network of managers in 2014 (nodes with highest leadership rank are also top managers)

Another interesting information can be derived from Figure 6, which shows the social network of managers in 2014. The shape of the graph is rather hierarchical, contrary to the any other role networks that we derived. Moreover, this structure closely matches with the actual role (and acknowledged influence) of managers in the company, which can be seen as an indication that, at least as far as managers are concerned, the leadership rank measure defined in this paper does have a correspondence with the company organizational chart. Note that only two women (the pink round nodes) appear in Figure 6, while a few other are isolated nodes, not visible in the Figure. Finally, interesting results arise from topic analysis by gender and role, which we forcefully omit for the sake of space.

## CONCLUDING REMARKS

In this paper we presented a set of measures and a platform to monitor and rank users' activity in an enterprise social network. Specifically, the platform is aimed at supporting the emergence (and recognition) of women leadership in organizations. Due to space restrictions, this paper was more concerned with describing the algorithms and main features of the platform than analyzing the outcomes of a case study on a large enterprise network, that we refer to future publications.

ACKNOWLEDGMENT

Fiordaliso has been funded by Regione Lazio under the FERS "*Insieme per Vincere*" program.

REFERENCES

[1] S. Grimsley Informal Leadership: Definition and Explanation, Retrieved from: http://study.com/academy/lesson/informal-leadership-definitionlesson-quiz.html on June, 2015.

[2] M. Neubert, and S. Taggar Pathways to informal leadership: The moderating role of gender on the relationship of individual differences and team member network centrality to informal leadership emergence. The Leadership Quarterly, Vol 15(2), Apr 2004, 175-194. http://dx.doi.org/10.1016/j.leaqua.2004.02.006

[3] A. H. Eagly. And S. J. Karau (2002) Role congruity theory of prejudice toward female leaders. Psychological Review, 109, 573–598

[4] Raina A. Brands, Jochen I. Menges, Martin Kilduff The Leader-in-Social-Network Schema: Perceptions of Network Structure Affect Gendered Attributions of Charisma, in Organization Science, ISSN 1047-7039, June 2015

[5] Peters, L., & O'Connor, E. (2001). Informal leadership support: An often overlooked competitive advantage. Physician Executive, 27(3), 35-35

[6] CEB analysis: Executive guidance 2014. The Rise of network leader, retrieved from http://www.executiveboard.com/exbd-resources/pdf/executive-guidance/eg2014-annual-final.pdf?cn=pdf on June 2015

[7] Quinn, R. E. & Spreitzer, G. M. 1997. The road to empowerment: Seven questions every leader should consider. Organizational Dynamics, 26(2): 37–49.

[8] Page, L., Brin, S., Motwani, R., and Winograd, T. (1999). The pagerank citation ranking: Bringing order to the web.

[9] Huang, B., Yu, G., and Karimi, H. R. (2014). The finding and dynamic detection of opinion leaders in social network. *Mathematical Problems in Engineering*, 2014.

[10] Lü, L., Zhang, Y.-C., Yeung, C. H., and Zhou, T. (2011). Leaders in social networks, the delicious case. *PloS one*, 6(6):e21202.

[11] Khorasgani, R. R., Chen, J., and Zaïane, O. R. (2010). Top leaders community detection approach in information networks. In *4th SNA-KDD Workshop on Social Network Mining and Analysis*. Citeseer.

[12] Pedroche, F., Moreno, F., González, A., and Valencia, A. (2013). Leadership groups on social network sites based on personalized pagerank. *Mathematical and Computer Modelling*, 57(7):1891–1896.

[13] Cao, J., Gao, H., Li, L. E., and Friedman, B. (2013). Enterprise social network analysis and modeling: A tale of two graphs. In *INFOCOM, 2013 Proceedings IEEE*, pages 2382–2390. IEEE

[14] Szell, M. and Thurner, S. (2013). How women organize social networks different from men. *Scientific Reports*, 3.

[15] A. Halu, R. Mondragon, P. Panzarasa and G. Bianconi: Multiplex Page Rank arXiv:1306.3576v3 [physics.soc-ph] 23 Sept. 2013

[16] D. Nemirovsky and K. Avrachenkov Proceedings of The Seventeenth Text REtrieval Conference, TREC 2008, Gaithersburg, Maryland, USA, November 18-21, 2008

[17] Fortunato S, Flammini A (2007) Random walks on directed networks: the case of PageRank. Int. Jour. Bifurcation and Chaos 17: 2343-2353

[18] Blei, D. M., Ng, A. Y., and Jordan, M. I. (2003). Latent dirichlet allocation. *the Journal of machine Learning research*, 3:993–1022

[19] R. Navigli and S. P. Ponzetto, "BabelNet: The automatic construction, evaluation and application of a wide-coveragemultilingual semantic network," Artificial Intelligence, vol.193, pp. 217–250, 2012

[20] Samudrala R.,Moult J.:A Graph-theoretic Algorithm for comparative Modeling of Protein Structure; J.Mol. Biol. (1998); vol 279; pp.